\documentclass[a4paper,fleqn,usenatbib]{mnras}

\usepackage[T1]{fontenc}
\usepackage{ae,aecompl}

\usepackage{graphicx}	
\usepackage{amsmath}	
\usepackage{amssymb}	

\title[Large polar caps of magnetars]{Large polar caps for twisted magnetosphere of magnetars}

\author[H. Tong]{H. Tong$^{1}$\thanks{E-mail: htong\_2005@163.com}
\\
$^{1}$School of Physics and Electronic Engineering, Guangzhou University, Guangzhou 510006, China
}

\date{Accepted XXX. Received YYY; in original form ZZZ}

\pubyear{2015}

\begin{document}
\label{firstpage}
\pagerange{\pageref{firstpage}--\pageref{lastpage}}
\maketitle

\begin{abstract}
The magnetic field of magnetars may be twisted compared with that of normal pulsars. Previous works mainly discussed magnetic energy release in the closed field line regions of magnetars. For a twisted magnetic field, the field lines will inflate in the radial direction. Similar to normal pulsars, the idea of light cylinder radius is introduced. More field lines will cross the light cylinder and become open for a twisted magnetic field. Therefore, magnetars may have a large polar cap, which may correspond to the hot spot during outburst. Particle flow in the open field line regions will result in the untwisting of the magnetic field. Magnetic energy release in the open field line regions can be calculated. The model calculations can catch the general trend of magnetar outburst: decreasing X-ray luminosity, shrinking hot spot etc. For magnetic energy release in the open field line regions, the geometry will be the same for different outburst in one magnetar.
\end{abstract}

\begin{keywords}
stars: magnetar -- stars: neutron -- pulsars: individual (XTE J1810$-$197; Swift J1822.3$-$1606)
\end{keywords}



\section{Introduction}

Magnetars are young and high magnetic field neutron stars (Duncan \& Thompson 1992; Kaspi \& Beloborodov 2017).
Observationally, they show various kinds of activities: bursts (including giant flares), outbursts, and variation of timing properties (Esposito et al. 2018). During the outburst, the magnetar persistent X-ray luminosity may be enhanced and decays subsequently. During the outburst decay phase, the magnetar may show a shrinking hot spot, appearance of radio emission and subsequent cessation, decreasing torque, softening X-ray spectra, and simple pulse profile etc (Coti Zelati et al. 2018; Esposito et al. 2018). The physical reason accounting for the magnetar activities may be that magnetars have twisted magnetic fields (Thompson et al. 2002). The twist may be bumped from interior in the neutron star. This internal strong magnetic field can also contribute to the magnetic energy release (Vigan\`{o} et al. 2013).

Magnetars typically have a rotational period about $P \sim 10 \, \rm s$. For such a long rotational period, the expected neutron star polar cap radius is very small, only about $50 \, \rm m$. Therefore, the closed field line regions of magnetars are mainly discussed previously (Thompson et al. 2002; Beloborodov 2009; Pavan et al. 2009). However, this argument is based on a pure dipole magnetic field. The magnetar's magnetic field may be highly twisted. For a twisted dipole magnetic field, it will inflate radially (Lynden-Bell \& Boily 1994; Wolfson 1995; Thompson et al. 2002). Similar to the case of normal pulsar, the light cylinder radius may be introduced to define the boundary between closed field lines and open field lines. Then for a twisted dipole field, more field lines will cross the light cylinder and become open. Therefore, magnetars may have large polar caps despite their long rotational period.

The particle flow in the open field line regions will be enhanced due to a large polar cap. This may also result in subsequent untwisting of the twisted magnetic field. Therefore, the open field line regions may provide an independent channel for the magnetic energy release in magnetars.

The numerical and analytical solutions of a twisted dipole field is presented in Section 2. For a twisted magnetosphere, the magnetar may have a large polar cap. This is demonstrated in Section 3. The consequences of a large polar of magnetars is presented in Section 4. Comparisons with observations are discussed in Section 5. The conclusion is given in Section 6.

\section{Axisymmetric force-free equilibrium}

\subsection{Basic equations}

The axisymmetric force-free equilibrium has been studied by many authors (Lynden-Bell \& Boily 1994; Wolfson 1995; Thompson et al. 2002; Gourgouliatos 2008; Pavan et al. 2009; Parfrey et al. 2012; Glampedakis et al. 2014; Akg\"{u}n et al. 2016; Kojima 2017). The symbolic system of Wolfson (1995) and Pavan et al. (2009) are adopted in the following.
The magnetosphere of pulsars and magnetars may be in the force-free equilibrium state. Assuming axisymmetric condition,  the magnetosphere is described by the Grad-Shafranov equation. In spherical coordinate, the Grad-Shafranov equation is:
\begin{equation}\label{eqn_GS}
\frac{\partial^2 A}{\partial r^2} + \frac{1-x^2}{r^2} \frac{\partial^2 A}{\partial x^2} + F(A) \frac{d F}{d A}=0,
\end{equation}
where $r$ is the dimensionless radial coordinate (in units of the neutron star radius), $x=\cos \theta$ ($\theta$ is the polar angle), $A=A(r,\theta)$ is the flux flunction, and $F(A)$ is a yet undetermined function (Wolfson 1995 and references therein).  The magnetic field is related to the flux function as (Wolfson 1995; Pavan et al. 2009)
\begin{equation}\label{eqn_B}
\bold{B} = \frac{1}{r \sin\theta} \left[  \frac{1}{r} \frac{\partial A}{\partial \theta}\hat{r} -\frac{\partial A}{\partial r} \hat{\theta} +F(A) \hat{\phi} \right].
\end{equation}
If the function $F(A)$ is in the form $F(A) = \lambda A^{1+1/n}$ ($\lambda$ and $n$ are numerical parameters), then equation (\ref{eqn_GS}) can be solved by separation of variables. The flux function will have the following form
\begin{equation}\label{eqn_flux}
A= r^{-n} f(x).
\end{equation}
From equation (\ref{eqn_flux}), the parameter $n$ reflects the radial dependence of the flux function. From equation (\ref{eqn_B}), the radial dependence of the magnetic field will be $B(r) \propto r^{-(2+n)}$. In general, the radial dependence of the magnetic field is different from the magnetic dipole field. The undetermined function $f(x)$ can be view as the dimensionless flux function. By the separation of variables, equation (\ref{eqn_GS}) is reduced to an ordinary differential equation
\begin{equation}\label{eqn_ODE}
(1-x^2) f^{''}(x) + n(n+1) f(x) + \lambda^2 \left(1+\frac{1}{n} \right) f^{1+2/n}(x)=0.
\end{equation}

When $\lambda=0$ and $n=1$, this corresponds to the magnetic dipole field. When $\lambda$ is different from zero, this means the presence of toroidal field, and the magnetic field is twisted compared with the dipole case. The polar axis should be a field line, this gives the boundary condition of equation (\ref{eqn_ODE}): $f(\pm 1) =0$. The boundary value requires that there is an eigenvalue of $n$ for a given $\lambda$ in equation (\ref{eqn_ODE}). For a twisted dipole field, the flux conservation gives the first initial condition: $f(0)=1$. Symmetry about the equational plane gives the second initial condition: $f^{'}(0)=0$. Starting from the two initial conditions, and considering the boundary condition requirement, equation (\ref{eqn_ODE}) can be solved numerically. For consideration near the polar cap regions, it is found that there are analytical solutions to equation (\ref{eqn_ODE}).

\subsection{Analytical solutions}

\begin{figure}
\centering
\includegraphics[width=0.45\textwidth]{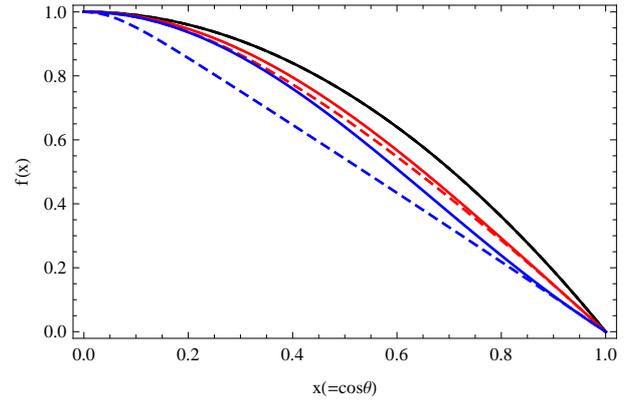}
\caption{Dimensionless magnetic flux as a function of polar angle. The solid lines are analytical approximations. The dashed lines are numerical calculations. The black, red and blue colors are for values of $n=1, 0.5, 0.1$, respectively. }
\label{gmagneticflux}
\end{figure}

For $\lambda=0$ and $n=1$, the solution of equation (\ref{eqn_ODE}) is the magnetic dipole field: $f_0(x) =1-x^2$. For an decreasing $n$ from $n=1$ to $n=0$, the magnetic field evolves from the dipole configuration to the split monopole. During this process, $\lambda$ will also be nonzero. For values of $1-n \ll 1$, an expansion around $\lambda=0$ and $n=1$ may be made. Following the treatment of Pavan et al. (2009), denote $\Delta n=n-1$, $f(x) = f_0(x) +f_1(x) \Delta n$, $\lambda^2 =\kappa \Delta n$. The expansion is made for
$\lambda^2$ because it is $\lambda^2$ which appeared in equation (\ref{eqn_ODE}). The calculation is straightforward (Pavan et al. 2009). It is found that $\lambda^2 =(35/16) (1-n)$, or
\begin{equation}\label{eqn_lambda}
\lambda = \sqrt{\frac{35}{16} (1-n)},
\end{equation}
where the positive root of $\lambda^2$ is chosen, which means the field line in the southern hemisphere of the neutron star will twist eastward (compared with field lines in the northern hemisphere).
The dimensionless flux function is
\begin{equation}\label{eqn_fapp}
f(x) =f_0(x) \left [  1- \frac{22-5x^2}{32}x^2 (1-n) \right ].
\end{equation}
Figure \ref{gmagneticflux} shows the comparison of analytical approximation with the numerical solution of equation (\ref{eqn_ODE}).
For $n$ not so small, the analytical solution is roughly consistent with the numerical calculations.

\begin{figure}
\centering
\includegraphics[width=0.45\textwidth]{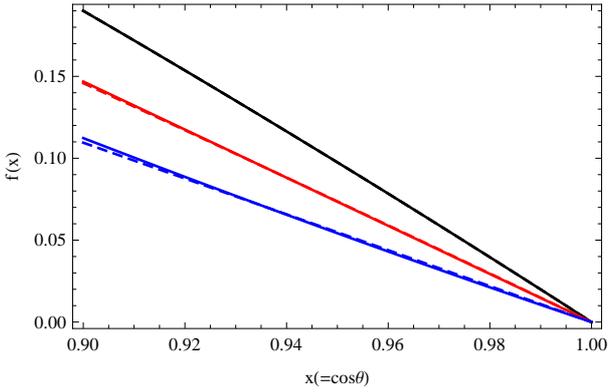}
\caption{Same as figure \ref{gmagneticflux}, for magnetic flux near the polar cap region.}
\label{gmagneticflux_polarcap}
\end{figure}

For the polar cap region, the polar angle is relatively small, e.g., $\theta \le 0.3$, and $x=\cos\theta > 0.9$. This will corresponds to a hot spot of $3 \, \rm km$ if observed\footnote{When using the Comptonized blackbody or other spectral models, the seed blackbody can come from either a hot spot or the whole neutron star surface (Enoto et al. 2017). Only when the blackbody radius is about several kilometers, it is thought to be from a hot spot. A clear example will be like that of XTE J1810$-$197, where both a hot spot component and a cold surface component can be seen (Alford \& Halpern 2016).}. In this case, the analytical solution can match the numerical solutions quite well, even for $n$ as small as $n=0.1$, see figure \ref{gmagneticflux_polarcap}. This feature of equation (\ref{eqn_fapp}) ensures that we can use the analytical solutions when calculating the geometry of the polar cap regions. Once the magnetic flux is obtained, the magnetic field, shear, current density, and magnetic energy can be calculated straightforward. In the following, we will try to provide analytical approximations when possible. The corresponding numerical results will also be shown for comparison.

\subsection{Maximum twist of magnetic field lines}

\begin{figure}
\centering
\includegraphics[width=0.45\textwidth]{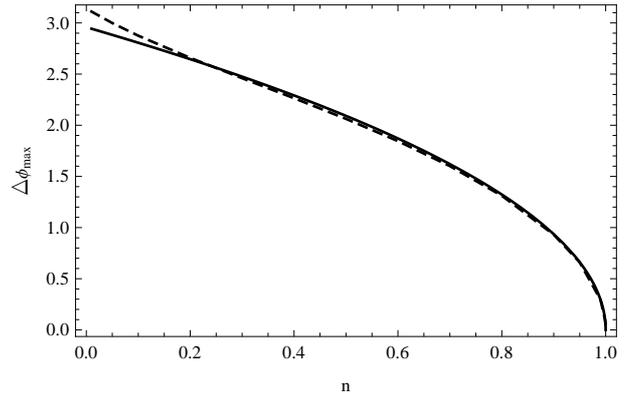}
\caption{Maximum twist as a function of the parameter $n$. The solid line is the analytical approximation. The dashed line is the numerical calculation.}
\label{gtwist}
\end{figure}

In the above self-similar solutions, the footpoints of the magnetic field lines will move not only in the latitudinal direction but also in the longitudinal direction (Lynden-Bell \& Boily 1994; Wolfson 1995). The longitudinal movement will result in the twist of the magnetic field lines. For a specific field line starts from $x_1= \cos\theta_1$ in the northern hemisphere and ends at $x_2 =\cos\theta_2$ in the southern hemisphere, the angular shear is (Wolfson 1995)
\begin{equation}
\Delta \phi =\frac{\lambda}{n} \int_{x_2}^{x_1} \frac{f^{1/n}(x)}{1-x^2} {\rm d}x,
\end{equation}
where during the definition of the twist a minus sign is absorbed. According to the definition of the twist, the field line starts at the north pole and ends at the south pole will have the maximum twist. Therefore, the maximum twist for a given equilibrium configuration is
(Thompson et al. 2002)
\begin{equation}\label{eqn_maximum_twist_def}
\Delta \phi_{\rm max} = \frac{2\lambda}{n} \int_{0}^{1} \frac{f^{1/n}(x)}{1-x^2} {\rm d}x.
\end{equation}
For $n\approx 1$ and $f(x) \approx 1-x^2$, the maximum twist is
\begin{equation}\label{eqn_maximum_twist}
\Delta \phi_{\rm max} =2\lambda.
\end{equation}
Figure \ref{gtwist} shows that the analytical approximation is quite accurate for a large range of parameter space.

For a specific equilibrium configuration, there is a one to one correspondence for the three parameters: $n$, $\lambda$, and $\Delta \phi_{\rm max}$. In some previous works, the maximum twist is used to characterize the state of the magnetic field lines.
However, during the definition of the maximum twist, it is assumed that the central magnetar does not rotate. Therefore, all the field lines are closed field lines and there is no open field lines. For real magnetars, they are rotating neutron stars. The field lines near the polar cap regions will become open field lines. Therefore, the definition of the maximum twist will be no longer valid.
In reality, as shown in below, magnetar can have large polar caps. This will further make the definition of the maximum twist inappropriate. Therefore, the parameter $n$ may be a better parameter to describe the state of a twisted magnetic dipole field. Physically, the parameter $n$ reflects the radial dependences of the magnetic field.

\section{Large polar caps}

In modeling the twisted magnetosphere of magnetars, the effect of rotation is neglected in several previous works (Thompson et al. 2002; Beloborodov 2009; Pavan et al. 2009; Glampedakis et al. 2014; Akg\"{u}n et al. 2016; Kojima 2017). Therefore, in these works, the central magnetar has no open field lines and no polar caps. For a typical rotational period of $10$ seconds, the polar cap region is very small for a dipolar magnetic field. This is the taken to be the reason why rotation is neglected in previous works. However, for a twisted dipole field, they tend to inflate in the radial direction. In analogy with normal pulsars, the open field lines are defined as those that pass through the light cylinder. Then more field lines will become open due to the inflation of field lines in the radial direction. Therefore, for a magnetar with twisted dipole magnetic field, it can have large polar caps despite its long pulsation period.

For a constant flux $A=r^{-n} f(x)={\rm constant}$ (equation (\ref{eqn_flux})), it corresponds to the projection of magnetic field lines in the $r-\theta$ plane. The dimension flux function $f(x)$ has the largest value at the equator (figure \ref{gmagneticflux}). Therefore, the maximum radial extension of the magnetic field line is also reached in the equatorial plane. The last closed field line can be defined as those with maximum radial extension equal to the light cylinder radius: $r_{\rm max}=R_{\rm lc} =P c/(2\pi)$, where $P$ the magnetar rotational period, and $c$ is the speed of light. The intersection of the last closed field line with the neutron star surface defines the boundary of the polar cap region. Since the flux function is a constant along a field line, then
\begin{equation}
\frac{f(0)}{R_{\rm  lc}^n} = \frac{f(x_{\rm pc})}{R^n},
\end{equation}
where $x_{\rm pc}$ is the angular radius of the polar cap, and $R$ is the neutron star radius. According to the conservation of magnetic flux $f(0)=1$, therefore the angular radius of the polar cap is determined by
\begin{equation}\label{eqn_flux_polarcap}
f(x_{\rm pc}) = \left ( \frac{R}{R_{\rm lc}} \right)^n.
\end{equation}
Note that $f(x_{\rm pc})$ is also the fraction of magnetic flux of the polar cap region. A typical magnetar is assumed to have rotational period of $10\,\rm s$, and a neutron star radius of $10\, \rm km$. For a magnetic dipole field, the parameter $n$ is $n=1$. Then the polar cap will have a fractional magnetic flux of $2\times 10^{-5}$. For a magnetic dipole field, the dimensionless flux function is: $f_0(x) =1-x^2 =1-\cos^2\theta =\sin^2\theta$. Then the angular radius of the polar cap is: $\sin\theta_{\rm pc} =\sqrt{R/R_{\rm lc}}$.
This is the result in normal pulsars (Goldreich-Julian 1969; Ruderman \& Sutherland 1975). Typically, the polar cap radius is only about $50$ meters for magnetars.

\begin{figure}
\centering
\includegraphics[width=0.45\textwidth]{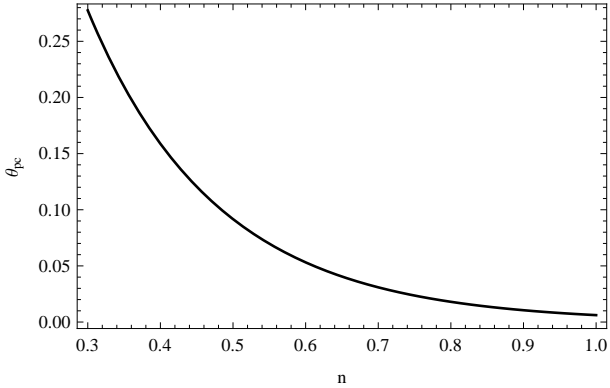}
\caption{Polar cap angular radius of a magnetar with twisted dipole field.}
\label{gthetapc_magnetar}
\end{figure}

From equation (\ref{eqn_flux_polarcap}), for a twisted dipole field with $n<1$, the polar cap will have larger fractional magnetic flux.
Furthermore, in equation (\ref{eqn_flux_polarcap}), only the value of $f(x)$ at the polar cap region is required. 
In the polar cap region with $x\approx 1$, the flux function can be further simplified: $f(x) =f_0(x) [1-((22-5x^2)/32) x^2 (1-n)] \approx f_0(x) [(15+17n)/32]$. The angular radius of the polar cap for a twisted dipole field is
\begin{equation}\label{eqn_polarcap}
\sin \theta_{\rm pc} =\sqrt{\frac{(R/R_{\rm lc})^n}{(15+17n)/32}}.
\end{equation}
Figure \ref{gthetapc_magnetar} shows the polar cap angular radius of a magnetar with twisted dipole field. For a angular radius of
$\theta_{\rm pc} =0.1$ or $\theta_{\rm pc} =0.2$, the corresponding polar cap radius is about $1\,\rm km$ or $2\,\rm km$. If the polar cap is heated by accelerated particles, a hot spot of $1-2 \,\rm km$ may be seen observationally.
Equation (\ref{eqn_polarcap}) reduces to the magnetic dipole case for $n=1$. If the magnetar's magnetic field is twisted during a star quake, during the untwisting process, the parameter $n$ evolves from $n<1$ to $n=1$. From figure \ref{gthetapc_magnetar}, during the untwisting process, the polar cap will decrease with time. Observationally, this may corresponds to the shrinking hot spot of magnetars during outbursts.

The footpoint of a twisted dipole field will move both in latitudinal direction and longitudinal direction (Lynden-Bell \& Boily 1994; Wolfson 1995).
The longitudinal movement will result in the twist of the magnetic field lines (see above). The latitudinal motion will also contribute to a larger polar cap of the magnetar. In the polar cap region, from the simplified flux function: $\sin^2\theta_{\rm pc} (n) (15+17n)/32 = \sin^2\theta_{\rm pc}({\rm dipole})$. For large twist ($n\ll 1$), the polar cap will be $\sqrt{2}$ times larger than the dipole case ($\theta_{\rm pc}(n) \approx \sqrt{2} \theta_{\rm pc}({\rm dipole})$) due to the latitudinal motion of the footpoint of the magnetic field lines. Therefore, the inflation of the magnetic field in the radial direction is the major reason for a large polar cap of magnetars.

\section{Particle flow in the open field line regions and untwisting}

\subsection{Magnetic free energy}

\begin{figure}
\centering
\includegraphics[width=0.45\textwidth]{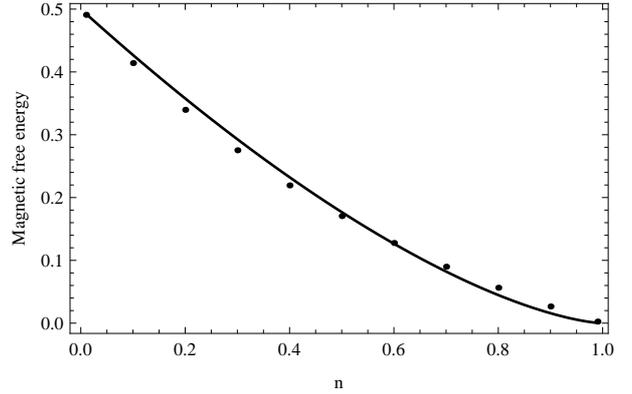}
\caption{Magnetic free energy of a twisted dipole field as a function of $n$. The black points are the numerical calculations. The solid line is the analytical fitting.}
\label{gEmf}
\end{figure}

For a twisted dipole field, the appearance of the toroidal component of the magnetic field means there are additional magnetic free energy compared with pure dipole case (Beloborodov 2009). For a twisted dipole field with radial dependence $B(r) \propto r^{-(2+n)}$, most of the magnetic energy are stored in the vicinity of the neutron star. Using the tensor virial theorem, the total magnetic energy is determined by the distribution of magnetic field at the surface of the neutron star\footnote{The numerical coefficient in equation (11) of Wolfson (1995) should be 3/2 instead of 2/3.} (Wolfson 1995)
\begin{equation}
E_{B} =\frac32 \int_{0}^{1} (B_{r}^2 -B_{\theta}^2 -B_{\phi}^2) {\rm d} x.
\end{equation}
The total magnetic energy is expressed in units of the total magnetic energy of the dipole magnetic field: $(1/12) B_{\rm p}^2 R^3$ (Thompson et al. 2002). Then the magnetic free energy is
\begin{equation}
E_{\rm mf} =E_{B}-1 =\frac32 \int_{0}^{1} (B_{r}^2 -B_{\theta}^2 -B_{\phi}^2) {\rm d}x -1.
\end{equation}
When there are open field line regions, the total magnetic free energy may be modified. The above treatment will be employed as the approximation to the real case.
The corresponding magnetic free energy as a function of parameter $n$ is shown in figure \ref{gEmf}. For a pure dipole field, $n=1$, the magnetic free energy is zero. For a split monopole, $n=0$, the magnetic free energy is $0.5$. Therefore, a not too bad analytical guess for the magnetic free energy is: $E_{\rm B, mf} =0.5(1-n)$. It is found that the following analytical expression fits the numerical results better
\begin{equation}
E_{\rm mf} =0.5 (1-n)^{1.5}.
\end{equation}
This analytical fitting will simplify relevant discussions and calculations.

\subsection{Particle flow in the open field line regions}

For normal pulsars, the particle flow, acceleration and radiation process mainly happen in the open field line regions (Goldreich \& Julian 1969; Ruderman \& Sutherland 1975; Cheng et al. 1986; Du et al. 2010). If the rotation of central magnetar is neglected from the starting point, all the field lines are closed field lines. Only particle acceleration and radiation in the closed field line regions can be considered (Thompson et al. 2002; Beloborodov 2009). Introducing the idea of light cylinder, as shown above, the central magnetar can have large polar caps. Then we can consider the particle flow, acceleration, and radiation in the open field line regions of magnetars, in analogy with that of normal pulsars.

For normal pulsars, the energy loss in the magnetosphere can be approximated roughly by the magnetic dipole radiation (Kou \& Tong 2015). The corresponding results can also be obtained by assuming a Goldreich-Julian current and a maximum acceleration potential for each flowing particle (Harding et al. 1999; Tong et al. 2013).  The maximum acceleration potential is the potential drop between the edge of the polar cap and the magnetic pole (Ruderman \& Sutherland 1975)
\begin{equation}\label{eqn_Vmax}
\Delta V_{\rm max} =\frac{\Omega R^2 B_{\rm p}}{2c} \sin^2\theta_{\rm pc},
\end{equation}
where $\Omega$ is the angular velocity of the central neutron star. For normal pulsars, with $\sin^2\theta_{\rm pc} = R/R_{\rm lc}$, the maximum acceleration potential is: $\Delta V_{\rm max} =\Omega^2 R^3 B_{\rm p} /(2c^2)$. The current through one polar cap is (Harding et al. 1999; Tong et al. 2013)
\begin{equation}\label{eqn_current}
I_{\rm pc} = \pi R_{\rm pc}^2 \rho_{\rm GJ} c,
\end{equation}
where $R_{\rm pc}$ is the polar cap radius, $\rho_{\rm GJ} =\Omega B_{\rm p}/(2\pi c)$ is the so-called Goldreich-Julian charge density (Goldreich \& Julian 1969).
Then the energy loss rate due to the particle flow is (i.e. particle luminosity, which is the energy carried by the accelerated particles and associated electromagnetic fields)
\begin{equation}\label{eqn_Edot_dipole}
\dot{E}_{\rm p, dipole} = 2 I_{\rm pc} \Delta V_{\rm max} = \frac{\Omega^4 R^6 B_{\rm p}^2 }{2c^3}.
\end{equation}
This result is the similar to that of magnetic dipole radiation, except for a different numerical factor.

For a twisted dipole field, the particle flow in the open field line regions will be enhanced due a large polar cap. For a larger polar cap, the corresponding maximum acceleration potential and polar cap current will be larger. This will result in a higher energy loss rate due to the particle flow. This enhanced particle flow is due to the twist of magnetic field lines. The particle acceleration and flow in the open field line regions will consume the magnetic free energy and results in the untwisting of the magnetic field lines. Similar to equation (\ref{eqn_Edot_dipole}), the corresponding particle luminosity for a twisted dipole field is
\begin{equation}\label{eqn_Edottwist}
\dot{E}_{\rm p, twist} = \frac{\Omega^2 R^4 B_{\rm p}^2}{2c} \sin^4\theta_{\rm pc},
\end{equation}
where $\theta_{\rm pc} \approx \sin \theta_{\rm pc}$ is used for small values of $\theta_{\rm pc}$ (i.e. polar cap regions).
As there may be multipole fields near the magnetar surface (Tiengo et al. 2013), a significant amount of the particle flow may be trapped by the multipole field.
By colliding with the neutron star surface, this particle luminosity may result in a hot spot on the neutron star and be converted into
X-ray luminosity of the magnetar. The physics may be similar to that in the closed field line regions (Beloborodov 2009). By colliding with the neutron star surface, a high conversion efficiency from the particle luminosity to the X-ray luminosity may be resulted. For a smaller conversion efficiency, it will only affect the normalization of the X-ray flux. The decaying pattern is the same. By chosing a higher magnetic field, a higher X-ray luminosity can be obtained again.

For normal pulsars, the particle acceleration and flow is still an unsolved question (Zhang et al. 2000; Kou \& Tong 2015).
The corresponding acceleration potential in various gap models is different from the maximum acceleration potential.
The flowing particle density can also deviates from the Goldreich-Julian density. The above treatment can be viewed as the strong particle flow case.
When the twisted dipole field is relaxed back to the pure dipole case, the polar cap will return  back to the dipole case. The corresponding particle luminosity (equation(\ref{eqn_Edottwist})) will also return back the dipole case. In this way, the magnetosphere of magnetars and normal pulsars may be unified together. This is merit of the above treatment.

\subsection{Untwisting}

According to energy conservation, the untwisting of the twisted dipole field is governed by the following equation
\begin{equation}\label{eqn_untwisting}
\frac{{\rm d} E_{\rm mf}}{{\rm d}t} = -\dot{E}_{\rm p,twist}.
\end{equation}
It should be noted that the magnetic free energy is expressed in units of the magnetic energy of a pure dipole field.

The particle luminosity should be of magnetic origin (instead of rotational origin, Lyutikov 2013). This is one basic assumption of our model. Justification of equation (\ref{eqn_untwisting}) from the energy conservation point of view is presented in the appendix. Different modeling of the particle luminosity will only result in quantitative differences.

\section{Discussion}

Observationally, the magnetar outbursts show a variety of changes: flux decay, shrinking hot spot, decreasing temperature, softening spectra, simpler pulse profile etc. The outburst may contain magnetic energy release from both the neutron star crust (Vigan\`{o} et al. 2013) or the magnetosphere (Thompson et al. 2002). Previous works considered the energy release in the closed field line regions (Beloborodov 2009). As shown above, the open field line regions may also contribute to the magnetic energy release. In below, the outburst of the first transient magnetar XTE J1810$-$197 is taken as an example. We will show to what degree the magnetic energy release in the open field line regions can explain the observations. 

\subsection{Decreasing X-ray luminosity}

\begin{figure}
\centering
\includegraphics[width=0.45\textwidth]{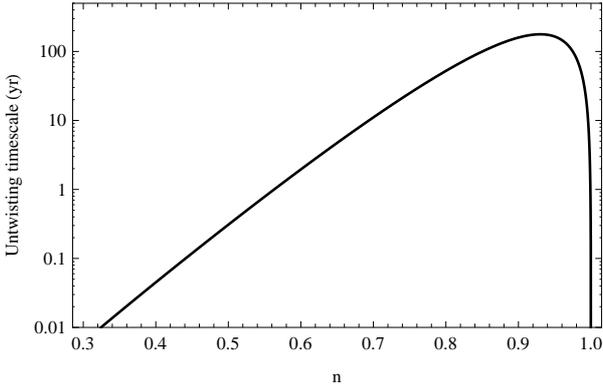}
\caption{Magnetic field untwisting timescale as a function of $n$. The untwisting timescale peaks at about $n=0.93$.}
\label{gtau}
\end{figure}

The total energy released by the hot spot in XTE J1810$-$197 is about $4\times 10^{42} \ \rm erg$ (Gotthelf \& Halpern 2007). The peak flux is about $10^{35} \ \rm erg \ s^{-1}$, and decays exponentially with a time constant about $200$ days (Gotthelf \& Halpern 2007). Later more observations showed a transition from three to two blackbody spectrum (Alford \& Halpern 2016). The hot spot component may be dominated by the energy release in the open field lines regions. For a typical parameter $n=0.5$, the total magnetic free energy is about $E_{\rm mf} \approx 10^{44} \ \rm B_{14}^2 \ erg$, the particle luminosity is about $\dot{E}_{\rm p, twist} \approx 10^{37} B_{14}^2\ \rm erg \ s^{-1}$. From equation (\ref{eqn_untwisting}), the magnetic energy decays with a typical timescale (i.e. untwisting timescale)
\begin{equation}\label{eqn_tau}
\tau(n) \equiv E_{\rm mf}/\dot{E}_{\rm p, twist} \sim 0.3 \rm \ yr \ (\text{for \ n=0.5}).
\end{equation}
Figure \ref{gtau} shows the untwisting timescale as a function of $n$. The untwisting time scale is not a monotonic function of the parameter $n$. It peaks at about $n=0.93$, which corresponds to a maximum twist about $\Delta \phi_{\rm max} \approx 0.8$. For large twist (i.e. small $n$), the magnetic free energy is approximately a constant. However, the particle luminosity decreases with increasing $n$. Therefore, the untwisting timescale increases with $n$ for large twist. For the small twist case, the magnetic free energy decreases as the field line untwists. At the same time, the particle luminosity is almost a constant. Therefore, the untwisting timescale decreases as the field lines untwists in the small twist case. This part is consistent with the small twist approximation in magnetar closed field line regions (Beloborodov 2009).
The over all behavior of the untwisting timescale is determined by the magnetic free energy and particle luminosity. Both of these quantities depends on the parameter $n$ in a non-linear way. This may explain the general behavior of the untwisting timescale.

The untwisting timescale in equation (\ref{eqn_tau}) and Figure \ref{gtau} represent a typical time interval for the twisted magnetosphere to lose a significant portion of its magnetic free energy. The total time for a twisted magnetosphere to become untwisted can be obtained by integrating equation (\ref{eqn_untwisting}). From Figure \ref{gtau}, the total time will be dominated by the maximum value of $\tau(n)$. Numerical calculations showed that the total untwisting time can be as long as\footnote{We note that the total untwisting timescale is the same order as the twist accumulation timescale found by Akg\"{u}n et al. (2017).} $600\,\rm yr$, and goes to zero as $n$ approaches $n=1$. However, this complete untwisting process will be too long to be observed. Furthermore, years after the outburst, the corresponding particle luminosity will be too low to be detected (detailed in below). At the same time, it may also be intervened by another outburst.

From Figure \ref{gtau}, we may get a dichotomy between the magnetosphere of magnetars and the magnetosphere of high magnetic field pulsars.
\begin{itemize}
   \item If a magnetar is induced to show outburst, e.g. by starquake, with initial parameter $n<0.93$ (or $\Delta \phi_{\rm max}> 0.8$), then its untwisting timescale will increase with time during the untwisting process. Years after the outburst, it will keep a magnetospheric configuration about $n\approx 0.93$, until the next outburst. The magnetic energy release rate is about $10^{33}\, B_{14}^2 \ \rm erg \ s^{-1}$. It may contribute to the quiescent luminosity of magnetars. Observationally, e.g. for XTE J1810$-$197, the quiescent luminosity contains a warm component plus a cold component from the whole neutron star surface (Alford \& Halpern 2016). The warm component may due to continued magnetic energy release during the quiescent state.
   \item If a neutron star is a high magnetic field pulsar initially, and a starquake twists it magnetosphere to $n>0.93$ (or $\Delta \phi_{\rm max} <0.8$), then its untwisting timescale will decrease with time. After sometime, it will return to a magnetosphere without twist. In this case, the maximum energy release rate due to the untwisting magnetosphere is $10^{33} \, B_{14}^2 \ \rm erg \ s^{-1}$. For young high magnetic field pulsars, their rotational energy loss rate will be significantly larger than this value. Therefore, the magnetic field powered activities in this case will be insignificant. Only when the initial $n$ parameter is smaller than $0.93$ (maximum twist larger than $0.8$), the magnetic activities may be significant. In this case, the untwisting behavior will be similar to that of magnetars. Up to now, two high magnetic field pulsar showed magnetar-like activities (Gavriil et al. 2008; Archibald et al. 2016; Gogus et al. 2016).
 \end{itemize}

\begin{figure}
\centering
\includegraphics[width=0.45\textwidth]{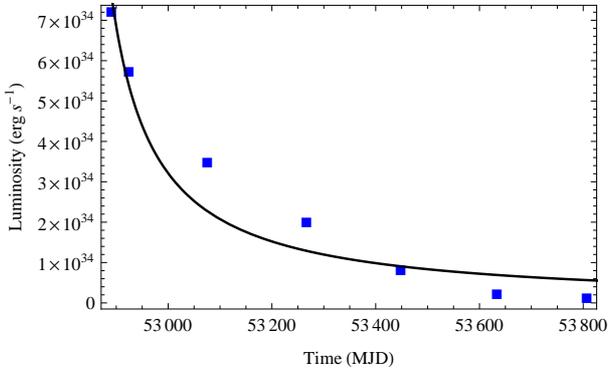}
\caption{Luminosity of the hot spot component of XTE J1810$-$197. The blue squares are observations (Alford \& Halpern 2016), the solid line is the model calculation.}
\label{glum}
\end{figure}

\begin{figure}
\centering
\includegraphics[width=0.45\textwidth]{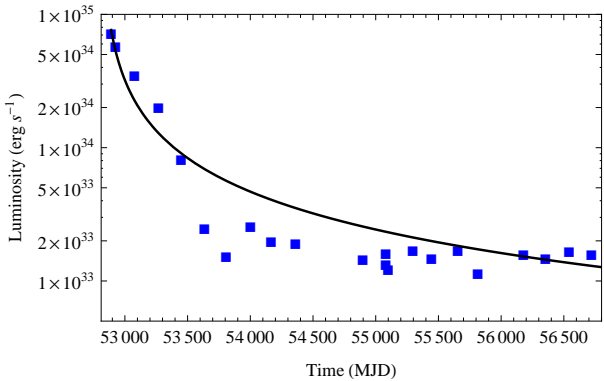}
\caption{Long term luminosity decay of XTE J1810$-$197. Similar to Figure \ref{glum}, the blue squares are observations (Alford \& Halpern 2016), the solid line is the model calculation.}
\label{glumlong}
\end{figure}

By solving Equation (\ref{eqn_untwisting}), the theoretical flux decay can be compared with the hot spot of XTE J1810$-$197.
From the total released energy and peak flux during the outburst, a surface magnetic field at the magnetic pole is chosen as $B_{14} =0.17$. This will be the polar dipole magnetic field strength when there is no twist. This value of magnetic field is for an energy conversion efficiency of unity. For a significantly smaller conversion efficiency (e.g., similar to that of normal pulsar $\sim 10^{-3}$), by chosing a higher magnetic field (e.g., 30 times higher), a high X-ray luminosity can be obtained again. The flux evolution time scale and evolution pattern is not affected by the energy conversion efficiency. For a parameter\footnote{At an earlier time, e.g. onset of the outburst, the parameter $n$ should be smaller.} $n=0.58$ at the time of the first XMM-Newton observation of XTE J1810$-$197 (MJD 52890.6), Figure \ref{glum} shows the hot spot luminosity of XTE J1810$-$197. Similar to Figure 4 in Alford \& Halpern (2016), the luminosities here are bolometric luminosities. The model calculations can catch the general trend of hot spot luminosity decay. At later time, the model calculation seems to over estimate the hot spot luminosity. However, at later time, the hot spot transforms to a warm-hot component (Alford \& Halpern 2016). Figure \ref{glumlong} shows the long term luminosity decay of XTE J1810$-$197. The first seven observational data in Figure \ref{glumlong} are the same as that in Figure \ref{glum}. The later observational data points in Figure \ref{glumlong} are the warm-hot component in Alford \& Halpern (2016). The theoretical long term luminosity is about $10^{33} \, \rm erg \ s^{-1}$. It can explain the warm-hot component of XTE J1810-197 during the quiescent state.

\subsection{Shrinking hot spot and temperature}

Observationally, XTE J1810$-$197 shows a transition from three blackbody spectra to two blackbody spectra (Alford \& Halpern 2016). It is possible that the central neutron star has a hot spot. The hot spot has a temperature gradient with the rest of the neutron star surface.
The magnetic energy release deposited onto the polar cap region may be diffused to the rest part of the neutron star surface.
Observationally, in order to explain the X-ray spectra of XTE J1810$-$197, a large neuron star radius is required $R\approx 30 \, \rm km$ (Alford \& Halpern 2016). This is too large to be modeled by our model. Such a large radius may be due to inaccurate source distance etc (see Alford \& Halpern 2016 for more discussions). The temperature of the hot spot can also be calculated theoretically. During the untwisting process, the acceleration potential for each particles decreases. Then the hot spot spot temperature is also expected to decrease with time. The theoretical hot spot temperature is about $1\, \rm keV$. While the observed temperature is about $0.6\, \rm keV$ (Alford \& Halpern 2016). Considering that the heat in the hot spot will diffuse to the rest part of the neutron star, it is natural that the theoretical hot spot temperature (not considering the diffusion process) is higher than the observations.


\subsection{Geometry and spin-down torque}

Here we considered the untwisting due to particle flow in the open field line regions of a globally twisted dipole field. In reality, the magnetar may contain higher order multipole fields. The magnetic energy release may also occur in the closed field line regions (Beloborodov 2009) or in the neutron star crust (Vigan\`{o} et al. 2013). One merit of considering particle flow in the open field line regions is that the geometry of hot spot will always be the same. If the radio emission also originates from the large polar caps of the twisted dipole field, then its magnetic field geometry is also not expected to change significant, except for a different twist at the onset of different outbursts. The revival of the magnetar PSR J1622$-$4950 (Camilo et al. 2018) and XTE J1810$-$197 (Gotthelf et al. 2019) both may require the same magnetic field geometry with the previous outburst.

For a dipole magnetic field, the twist will result in a large polar cap, enhanced particle flow, stronger magnetic field at the light cylinder radius, and the presence of a strong toroidal field. All these aspects will contribute to a larger torque than the pure dipole case. This may explain why the required magnetic field of XTE J1810$-$197 is smaller than its characteristic magnetic field. During the untwisting process, a decreasing torque is expected. This is in general consistent with the timing observations of XTE J1810$-$197 (Camilo et al. 2016; Levin et al. 2019).

For a twisted dipole field, the light cylinder radius determines the boundary of the polar cap region. A large polar cap will result in a stronger particle flow. This strong particle flow may also result in the opening of the magnetic field lines at a smaller radius than the light cylinder radius (Harding et al. 1999; Tong et al. 2013). A smaller opening radius will again result in a larger polar cap. Therefore, the torque due to the particle flow in the case of twisted magnetic field should be treated in a self-consistent way. This may be the future works.


\subsection{Statistical properties of magnetar outbursts}

The outburst observations of magnetars are diverse and rich (Esposito et al. 2018; Coti Zelati et al. 2018). The above calculations for XTE J1810$-$197 is only one example. From Equation (\ref{eqn_Edottwist}), the particle luminosity is proportional to the square of the polar cap area. This may result in a X-ray luminosity proportional to the square of the hot spot area during the outburst decay of the magnetar: $L_{\rm x} \propto A^2$, where $A$ is the hot spot area. This relation is for the case of maximum acceleration potential. For a constant acceleration potential, the particle luminosity will be proportional to the polar cap area. This may result in a X-ray luminosity proportional to the hot spot area: $L_{\rm x} \propto A$. The real case may lie between these two cases. Therefore, a correlation between the X-ray luminosity and hot spot area will be: $L_{\rm x} \propto A^{\alpha}$, where the power law coefficient $1<\alpha<2$ is expected. Long term flux decay of the magnetar Swift J1822.3$-$1606 found a power coefficient $\alpha<2$ (Scholz et al. 2014).

As been discussed above, long after the outburst, the magnetar may enter into a quiescent state with $n=0.93$ and typical magnetic energy release rate $\sim 10^{33} \, B_{14}^2 \rm erg \ s^{-1}$. However, years after the outburst, the decay time scale is long enough that the magnetar may already been considered as in the quiescent state. Therefore, the corresponding theoretical quiescent luminosity will be $>10^{33} \, B_{14}^2 \rm erg \ s^{-1}$. A general correlation between magnetar quiescent luminosity and dipole magnetic field is: $L_{\rm x,q} \propto B_{\rm p}^2$. For the case of a constant acceleration potential, the correlation will be: $L_{\rm x,q} \propto B_{\rm p}$. Therefore, the power law index between one and two is expected for $L_{\rm x,q} \propto B_{\rm p}^{\beta}$, where $1<\beta<2$. Observationally, such a correlation is indeed found by Coti Zelati et al. (2018). However, when interpreting the observations, two cautions should be made: (i) Observationally, the quiescent luminosity may include both the magnetic energy release and the a cold component from the whole neutron star surface (e.g. XTE J1810$-$197, Alford \& Halpern 2016). (ii) The neutron star spindown torque may be significantly enhance compared with the dipole case. The characteristic dipole magnetic field may just be a measure of the total torque. It can be significantly larger than the true dipole magnetic field (Tong et al. 2013). Albeit with these two uncertainties, we think that the correlation between magnetar quiescent luminosity and characteristic magnetic field is another evidence that are consistent with our model.

\subsection{Comparison with previous works}

In Thompson et al. (2002), a globally twisted magnetosphere is considered. While in Beloborodov (2009), a locally twisted magnetic field is explored. There are no open field lines. In Beloborodov (2009), magnetic energy release in the closed field line regions is considered. In our model, the magnetar may have a large polar cap due the twist of the dipole magnetic field. Possible magnetic energy release in the open field line regions are considered. This will also result in untwisting of the magnetic field. Therefore, our model provides an independent channel for magnetic energy release in the case of magnetars. Magnetic energy release may be at work simultaneously in both the closed field line regions and open field line regions.

For magnetic energy release in the closed field line regions and the same magnetar, different outbursts may happen at different locations.
While for magnetic energy release in the open field line regions, different outburst will always happen at the polar cap regions. Therefore if magnetic energy release occurs in the open field line regions, the magnetar will exhibit the same geometry during different outbursts.
The recurrent outburst of XTE J1810$-$197 may help us to solve this problem observationally, at least for this source.

Glampedakis et al. (2014) demonstrated three kinds of twist in the neutron star magnetosphere: (i) Twist  in the closed field line regions. This may corresponds to magnetar magnetosphere with some localized twist. (ii) Twist in the open field line regions. This may mimic the magnetosphere of normal pulsars. (iii) Twist in both the open and closed field line regions. This may corresponds to a globally twisted magnetosphere of magnetars (Thompson et al. 2002; Pavan et al. 2009; and this work). There are many works on magnetar magnetosphere with localized twist (Gourgouliatos 2008; Beloborodov 2009; Fujisawa \& Kisaka 2014; Akg\"{u}n et al. 2016, 2017). The effect of relativity are also explored (Gourgouliatos \& Lynden-Bell 2008; Pili et al. 2015; Kojima 2017; Huang et al. 2018). Time dependent twisted magnetosphere and its effect on the pulsar spin-down torque are studied by Parfrey et al. (2012, 2013). Compared with previous works, we try to model the magnetar magnetosphere from a global twisted point of view. The effect of pulsar rotation is not considered in a self-consistent way. It is done similar to that of normal pulsars. By defining the light cylinder radius, the open field lines and closed field lines are classified. The consequence is a large polar cap for magnetars, in spite of its long pulsation period.

\section{Conclusion}

We found that magnetars may have large polar caps despite their long pulsation period. Possible magnetic energy release in the open field line regions are also explored. Previously, the polar cap of magnetars is thought to be very small due to their long rotational period ($P \sim 10\,\rm  s$). Magnetic energy release in closed field line region are mostly discussed. Considering that the magnetar's magnetic field may be twisted and a twisted magnetic field tend to inflate in the radial direction. This will result in a large polar cap for magnetars. The consequences of a large polar cap are enhanced particle flow in the open field line regions and untwisting of the magnetic field etc. Our model calculations can catch the general trend of magnetar outburst decay.

We modeled the magnetar magnetosphere from a globally twisted point of view. Localized multipole field (Tiengo et al. 2013) can not be model in our model. Possible magnetic energy release in the open field line regions are explored and applied to magnetar outbursts. Magnetar outbursts are of relative long timescale ($\sim$ years). Short timescale bursts and giant flares (Parfrey et al. 2012) are not the object of this work. The effect of magnetar rotation is considered mainly by introducing the light cylinder radius. The pulsar magnetosphere may be realized by solving the pulsar equation with dipole magnetic field boundary condition (Contopoulos et al. 1999). Therefore, the possible large polar cap of magnetars may be testified by future works solving the pulsar equation with the twisted dipole magnetic field as the boundary condition. Possible magnetic energy release in the open field line regions are modeled by assuming Goldreich-Julian particle density and a maximum acceleration potential. As in the case of normal pulsars, the particle density can by higher than the Goldreich-Julian density (which is actually the charge density). The physical acceleration potential may also deviated from the maximum acceleration potential (Kou \& Tong 2015 and references therein). Similar things may also happen in the case of magnetars. These may be the topics of future works.

\section*{Acknowledgements}

The authors would like to thank H. G. Wang for discussions and the referee for insightful suggestions.
H.Tong is supported by NSFC (11773008).

\appendix

\section{Justification of Equation (20)}

The presence of current and acceleration field means that the field will done work on the particles at an rate: $\bf{j}\cdot \bf{E}$.
After some manipulation, it can be proven that (Section 2.1 in Rybicki \& Lightman 1979):
\begin{equation}\label{eqn_Poynting}
\frac{d}{dt} (U_{\rm mech} + U_{\rm field}) = - \int \bf{S}\cdot d\bf{A},
\end{equation}
where $U_{\rm mech}$ is the total mechanical energy inside the volume, $U_{\rm field}$  is the total field energy inside the volume, $\bf{S}$ is the Poynting flux, and the integration is over an closed bounding surface. This equation states that: the total outflow Poynting luminosity is provided by the decrease in the total mechanical and field energy. Draw an imaginary sphere of radius $R_{\rm lc}$ around the neutron star. At the light cylinder radius, the corotational velocity is about the speed of light. Therefore, $\bf{E} \approx \bf{B}$ at the light cylinder. The total Poynting luminosity through this sphere is
\begin{equation}
S_{\rm tot} \sim \frac{c}{4\pi} B(R_{\rm lc})^2 4\pi R_{\rm lc}^2.
\end{equation}

The magnetic field decrease with as $r^{-(2+n)}$ for a twisted dipole field.
\begin{itemize}
  \item For $n=1$ (the pure dipole case), the total Poynting luminosity is $S_{\rm tot} \sim \Omega^4 R^6 B_{\rm p}^2/c^3$. It is the same as equation (\ref{eqn_Edot_dipole}), except for a different numerical factor. In this case, the toroidal field in the vicinity of the neutron star is very small and is caused by the rotation of the neutron star (Harding et al. 1999). The magnetic free energy is also very small compared with the neutron star rotational energy. Therefore, in the case of pure dipole, the total Poynting luminosity should be powered by the neutron star rotational energy.

  \item For $n\neq 1$ (twisted dipole case), the total Poynting luminosity is
\begin{equation}
S_{\rm tot} \sim \frac{\Omega^2 R^4 B_{\rm p}^2}{c} \left( \frac{R}{R_{\rm lc}} \right)^{2n}.
\end{equation}
Using equation (\ref{eqn_polarcap}), the polar cap angular radius is approximately: $\sin^2\theta_{\rm pc} \approx (R/R_{\rm lc})^n$.
Therefore, the total Poynting luminosity is
\begin{equation}
S_{\rm tot} \sim \frac{\Omega^2 R^4 B_{\rm p}^2}{c} \sin^4\theta_{\rm pc}.
\end{equation}
Again, the total Poynting luminosity is the same as equation (\ref{eqn_Edottwist}), except for a different numerical factor.
On the right hand of equation (\ref{eqn_Poynting}), the total rotational energy of the neutron star is $U_{\rm mech} \approx 6\times 10^{44} \, \rm erg$ (for magnetar XTE J1810$-$197, whose rotational period is $P=5.54 \, \rm s$, Pintore et al. (2019)). For a typical parameter $n=0.5$, the total magnetic free energy is $U_{\rm field} \approx 10^{44} B_{14}^2 \,\rm erg$. The magnetic free energy is comparable with the total rotational energy. Furthermore, for $n=0.5$, the typical Poynting luminosity is $S_{\rm tot} \sim 10^{37} B_{14}^2 \,\rm erg \, s^{-1}$. Observationally, the rotational energy loss rate is only about $6\times 10^{32} \, \rm erg \, s^{-1}$ (Pintore et al. 2019). During the outburst, the rotational energy loss rate will be several times higher. However, it is still much smaller than the total Poynting luminosity. Therefore, in the case of twisted dipole field, the total Poynting luminosity should be powered by the magnetic energy. This justifies equaton (\ref{eqn_untwisting}) in our paper.
\end{itemize}

The physical reason for the above differences is origin of toroidal magnetic field. The discharge in the closed field line regions and open field line regions are both driven by the magnetic twist (which will result in an electric current) (footnote 4 in Beloborodov 2009). In the case of normal pulsars, the toroidal field (although relatively small) is due to the neutron star rotation. Therefore, the particle luminosity is powered by the rotational energy. In the case of magnetars, the toroidal field is due to the magnetic twist, either global (our model) or local twist (Beloborodov 2009). Therefore, the particle luminosity is powered by the magnetic energy in the case of magnetars.







\bsp	
\label{lastpage}
\end{document}